\documentstyle[12pt]{article}
\begin{document}
\baselineskip=7mm
\newcommand{\TeV}{\,{\rm TeV}}
\newcommand{\GeV}{\,{\rm GeV}}
\newcommand{\MeV}{\,{\rm MeV}}
\newcommand{\keV}{\,{\rm keV}}
\newcommand{\eV}{\,{\rm eV}}
\newcommand{\be}{\begin{equation}}
\newcommand{\ee}{\end{equation}}
\newcommand{\bea}{\begin{eqnarray}}
\newcommand{\eea}{\end{eqnarray}}
\newcommand{\ba}{\begin{array}}
\newcommand{\ea}{\end{array}}
\newcommand{\bmat}{\left(\ba}
\newcommand{\emat}{\ea\right)}
\newcommand{\refs}[1]{(\ref{#1})}
\newcommand{\ler}{\stackrel{\scriptstyle <}{\scriptstyle\sim}}
\newcommand{\ger}{\stackrel{\scriptstyle >}{\scriptstyle\sim}}
\newcommand{\lag}{\langle}
\newcommand{\rag}{\rangle}
\begin{titlepage}
\title{Natural $\mu$-term with Peccei-Quinn Symmetry\\
\vspace{-3cm}
\hfill\parbox{3cm}{\normalsize IC/94/358 \\ November 1994}
\vspace{3cm}}
\author{E.~J.~Chun\thanks{Email address: chun@ictp.trieste.it}\\[0.2cm]
        International Centre for Theoretical Physics\\
        P.O.Box 586, 34100 Trieste,  Italy }
\date{}
\maketitle
\vspace{3cm}
\begin{abstract}
\baselineskip=7mm {\normalsize
The generalized Higgs mass term $NH_1H_2$ of the supersymmetric standard
model is used to  implement the Peccei-Quinn Symmetry to solve the
strong-CP problem.  Then supersymmetry breaking can generate the Higgs mass
parameter $\mu$ of order $m_{3/2}$ through soft-breaking parameters.
This kind of extension contains extra light fields of the axion supermultiplet
whose dominant coupling may  come from the supersymmetric
axion-Higgs-Higgs coupling.  We present a working example and
discuss the cosmological implications of the model.}
\end{abstract}

\thispagestyle{empty}
\end{titlepage}
\clearpage

\setcounter{page}{1}

The supersymmetric Higgs mass term $\mu H_1 H_2$ in the minimal supersymmetric
standard model (MSSM) brings a problem of naturalness, so called the
$\mu$-problem \cite{kn}.  The $\mu$-term together with soft supersymmetry
breaking terms drive electroweak symmetry breaking.  Therefore both
the parameter $\mu$ and soft-breaking parameters  are required to be at
or slightly above the electroweak scale.  The scale of soft-breaking parameters
(characterized by the gravitino mass $m_{3/2}$)
can be understood from the hidden-sector supersymmetry breaking
mechanism \cite{nil}. The question is why the $\mu$-parameter is  so small
compared to the other scale in the theory, e.g., the Planck scale $M_{Pl}$.
The $\mu$-term may well be generated dynamically due to supersymmetry
breaking.

One conventional way to explain the origin of the $\mu$-term is to
include a singlet $N$ under the standard model (SM) gauge group and
introduce a term $N H_1 H_2$ \cite{nsw}.
Then one can arrange for the vacuum expectation
value (VEV) of $N$ to get the proper value.
In this case one generically expects the presence of extra
light singlets. Very recently it has been
demonstrated that introducing an $U(1)$ gauge group can achieve the
generation of the $\mu$-term without producing light singlets \cite{hemp}.
This approach can make $N$ as heavy as $M_{Pl}$ but calls for  some light
colored fields in order to cancel the anomaly.
\bigskip

In this letter we stress that the generation of the $\mu$-term can be
made much more appealing if one uses the Peccei-Quinn (PQ) Symmetry
\cite{pq} instead of any other global or local symmetry.
Obviously one can combine the $\mu$-problem and the strong-CP problem
\cite{kim} in this way.  If one wants to understand the strong-CP problem
in terms of the PQ mechanism, it is quite natural to assign the presence
of the term $h NH_1H_2$ to the PQ symmetry.
Then one has to arrange certain Higgs
superpotential providing PQ symmetry breaking at the invisible scale
$M_{PQ} \sim 10^{10}-10^{12} \GeV$. In this case $\lag N \rag \sim M_{PQ}$
implies extreme fine-tuning of $h$.  But it is not necessary for $N$
to have such a large VEV as is commonly believed. If the VEV of $N$ vanishes
in the supersymmetric limit, soft supersymmetry breaking may  induce
nonvanishing VEV which is expected to be of order $m_{3/2}$.
The PQ symmetry is broken by some other fields which couple to $N$.
\bigskip

In the first attempt to relate the dynamical generation of the $\mu$-term
to the strong-CP problem, a non-renormalizable term like
$S^2 H_1 H_2/M_{Pl}$ was used \cite{kn}.
For this to provide the proper value of $\mu$, the PQ scale
($ \lag S \rag \sim M_{PQ}$) is necessarily of the same order as the hidden
sector supersymmetry breaking scale which is indeed inside the
above-mentioned window for the PQ scale.
However our prescription with the renormalizable term is in
fact irrelevant to the size of the PQ scale, which makes it viable even if
any dissipation mechanism of the axion energy density works to remove the
upper bound of the PQ scale \cite{bbs}.

Other approaches to  the $\mu$-problem in supergravity or
superstring theories have been investigated  in refs.~\cite{ckn}
with the PQ symmetry and  in refs.~\cite{gm} and \cite{zw} without it.
\bigskip

In the below we will illustrate that the VEV of $N$ is generated due to
soft-breaking parameters and the PQ symmetry breaking is driven by the VEV's of
some other fields (called $S$).  Only one light degree of freedom (that is,
the axion supermultiplet $\Phi$) is added to the MSSM.  The couplings of the
axion supermultiplet to the MSSM particles are well-fixed due to the nature of
the PQ symmetry.  The generalized $\mu$-term $NH_1H_2$ induces the
supersymmetric  axion-Higgs-Higgs vertex  $\Phi H_1 H_2$ which may provide
the dominant coupling of the axion and its superpartners.
The mass splitting inside the axion supermultiplet and its cosmological role
through the $\mu$-term coupling will be also discussed.
\bigskip

Our working example introduces five ``PQ fields'' $N, S'', S, S'$ and $Y$
which are singlets under the SM gauge group and their PQ charges are assinged
to be $X = (-2, 2, 1, -1, 0)$. Furthermore we impose the R-symmetry under which
the PQ fields carry the charges $R = (2,0,0,0,2)$.
The most general ``PQ superpotential'' invariant under both the PQ symmetry
and the R-symmetry is
\be \label{mo}
 W_{PQ} = M NS'' + f N S^2 + f' (SS'-M^{'2})Y
\ee
together with the term $hNH_1H_2$. Note that the renormalizable term $S''S'^2$
and the non-renormalizable term $H_1H_2S'^2/M_{Pl}$ are forbidden by the
R-symmetry, which is crucial for our discussion.
In the supersymmetric limit, the VEV's are given by
\bea \label{su}
 \lag SS' \rag &=& M^{'2} \nonumber\\
 \lag S'' \rag &=& -f \lag S \rag^2/M \nonumber  \\
 \lag Y \rag &=& 0 \nonumber\\
 \lag N \rag &=& 0 \;.
\eea
The Goldstone (axion)  mode is a linear combination of $S,S'$ and $S''$.
The PQ scale is given by $M_{PQ} = (\lag S \rag^2 + \lag S' \rag^2 + 4\lag S''
\rag^2)^{1/2}$.  One can see that all the other modes get masses
of order $M_{PQ}$.

Inclusion of soft supersymmetry breaking effect gives the desired feature.
The scalar potential should be completed with the soft-terms,
\bea
 V_{soft} &=& BMNS'' + fANS^2 + f'A'SS'Y - f'B'M^{'2}Y + \mbox{h.c.}
                    \nonumber\\
          &&  + m_N^2 |N|^2 + m_S^2 |S|^2 + m_{S'}^2 |S'|^2
            + m_{S''}^2 |S''|^2
\eea
where $A,B,A',B'$ and $m_i$ are soft-breaking parameters of order $m_{3/2}$.
Minimization of the full scalar potential yields the following changes from
eq.~\refs{su}:
\bea \label{so}
 \lag N \rag &=& {2f(A'-B')-f(1+r^2)(A-B) \over \xi(r+r^{-1})+4f^2r^2} \\
 \lag Y \rag &=& {2f^2r^{-1}(A-B)- f^{'-1}(\xi+4f^2r^{-1})(A'-B')
                        \over \xi(r+r^{-1})+4f^2r^2} \nonumber
\eea
where $r \equiv \lag S/S' \rag$ and $\xi \equiv M^2/M^{'2}$.
The value of $r$ depends on the parameters, which we do not show explicitly.
The VEV's of $S,S'$ and $S''$ get negligible changes of order
$m_{3/2}^2/M_{Pl}$.  We note that in the limit $f \rightarrow 0$, we have
$\lag N \rag \rightarrow 0$ and $\lag Y \rag \rightarrow -(A'-B')/f'$
as discussed in the connection with supersymmetric majoron models
\cite{mz}.
\bigskip

We have seen that the light degrees of freedom  contain the axion
supermultiplet in addition to the usual MSSM particles.
The couplings of the axion supermultiplet to the MSSM particles are just
supersymmetric counter parts of the conventional
Dine-Fischler-Srednicki-Zhitnitskii axion model \cite{dfsz}.
In addition,  the generalized $\mu$-term $hNH_1H_2$ implies the supersymmetric
axion-Higgs-Higgs coupling
\be
 W_{ahh} = X_N {\mu \over M_{PQ}} \Phi H_1H_2
\ee
where $X_N = -2$ and $\mu=h\lag N \rag$. It can play an important role
in the supersymmetric axion models.
The mass splitting among the axion and its supersymmetric particles (called
the saxion and the axino) can be also calculated in our model.
The axino mass is given by $m_{\tilde{a}} = f'\lag Y \rag$ and the saxion mass
squared is certain linear combination of $m^2_S, m^2_{S'}$ and $m^2_{S''}$.
Both of them get masses of order $m_{3/2}$ without accepting any
fine-tunning of the parameters.
\bigskip

We now turn to the cosmology of the model.  Since the parameter $\mu$ is
of order 100 GeV, the above axion-Higgs-Higgs vertex may give  the most
strongest interaction of the axion supermultiplet to the MSSM fields.
The axino and the saxion with masses of order $m_{3/2}$ have to
be unstable \cite{rtw}.  Then the axino (saxion) may decay into
a Higgsino and a Higgs (two Higgses).
A cosmological bound on the axino and the saxion mass comes from the standard
nucleosynthesis. Since the axino and the saxion decouple at very high
temperature, their relic densities overdominate the energy density of the
universe when they decay.  Then in order not to destroy the prediction of the
nucleosynthesis their lifetime should be  shorter than about 1 sec.
The axion-Higgs-Higgs vertex induces fast enough decay to satisfy this
constraint.
For the axino we have another constraint from the fact that its decay products
should contain at least one lightest supersymmetric particle (LSP) which
forms cold dark matter of the universe. In order to avoid the overclosure due
to the decay-produced LSP's, the axino decay should occur before the LSP
decouples.  Taking the typical decoupling temperature of the LSP at about
$T=m_{LSP}/20$, we get the bound on the axino mass
\be \label{b1}
  m_{\tilde{a}} \ger 190 \GeV  \left(m_{LSP} \over 20 \GeV\right)^2
                      \left(200\GeV \over \mu \right)^2
                       \left( M_{PQ} \over 10^{12} \GeV\right)^2 \,.
\ee
If $\mu$ is smaller than  the top quark mass,  the axino may decay into
a top and a light stop in which case one replace $\mu$ by the top quark mass.

The axino itself can be the LSP when it is lighter than a few keV providing
the underclosure energy density of the universe.  One may be able to find a
PQ superpotential which admits such a light axino \cite{gy}.
\bigskip

If one would like to have the $\mu$-parameter smaller than 100 GeV or the
larger $m_{LSP}$ , the above bound produces  a bit large amount of
splitting between the axino mass and the $\mu$-parameter.
In the presented model, the required splitting can be
obtained by tuning the parameters $f$ and $f'$.  However, better way to
overcome this constraint is to invoke inflation.
The decoupling temperature of the axino (saxion) may be higher
than the reheating temperature $T_R \sim 10^{10}$ GeV which is a
maximally allowed value to cure the gravitino problem
in supergravity models \cite{ekn}.
Then the primordial axino relics are diluted away.
We recall that the axino decoupling temperature is determined by the
annihilation of an axino and a gluino into two quarks \cite{rtw}, which gives
the decoupling temperature;
\be
  T_D \sim 10^{11} \GeV \left( M_{PQ} \over 10^{12} \GeV \right)^2
                   \left(0.1   \over \alpha_c \right)^3 \,.
\ee
After inflation, the universe can produce the axino population
through inequilibrium process of the inverse annihilation.
The ratio of the regenerated number density to the entropy density
is given by \cite{ekn}
\be
  Y \sim 7 \times 10^{-5} \left( 10^{12} \GeV \over M_{PQ} \right)^2
                     \left( T_R \over 10^{10} \GeV \right) \,.
\ee
The regenerated number of axinos should be sufficiently reduced in order to
avoid overclosure due to the decay-products.  It gives the upper bound
on the reheating temperature;
\be \label{b2}
  T_R \ler 1.4 \times 10^7 \GeV \left(M_{PQ} \over 10^{12} \GeV \right)^2
                    \left( 20 \GeV \over m_{LSP} \right) \;.
\ee
Therefore, either the lower bound on the axino mass \refs{b1} or the
upper bound on the reheating temperature \refs{b2} has to be satisfied
in the inflationary universe.
\bigskip

In conclusion, we have illustrated a mechanism  of generating  the  $\mu$-term
in the axionic extension of the MSSM.  Soft supersymmetry breaking parameters
induce the natural value of order $m_{3/2}$ for the  parameter $\mu$.
This scheme is precisely the supresymmetric version of the
Dine-Fischler-Srednicki-Zhitnitskii axion model which adds the light fields
from the axion supermultiplet to the MSSM.
The supersymmetrized axion-Higgs-Higgs interaction induced from the extended
Higgs mass term $NH_1H_2$ can provide the main decay mode of the superpartners
of the axion. From this we have drawed the bound on the axino mass or
on the reheating termperature in the inflationary universe.
\bigskip

{\bf Acknowledgement}: The author  would like to thank Professor Abdus Salam,
the International Atomic Energy Agency and UNESCO for hospitality at the
International Centre for Theoretical Physics, Trieste.


\newpage
\begin{thebibliography}{99}
\bibitem{kn}  J.E. Kim and H.P. Nilles, Phys.~Lett.~B138 (1984) 150.
\bibitem{nil}  H.P. Nilles, Phys.~Rep.~110 (1984) 1.
\bibitem{nsw}  H.P. Nilles, M. Srednicki and D. Wyler, Phys.~Lett.~B120 (1983)
               346; J.M. Frere, D.R.T. Jones and S. Raby, Nucl.~Phys.~B222
              (1983)11; J.P. Derendinger and C. Savoy, Nucl.~Phys.~B237 (1984)
              307; L.E. Ibanez and J. Mas, Nucl.~Phys.~B286 (1987) 107;
              J. Ellis, J.F. Gunion, H.E. Haber, L. Roszkowski and F. Zwirner,
              Phys.~Rev.~D39 (1989) 844; M. Drees, Int.~J.~Mod.~Phys.~A4
             (1993) 288.
\bibitem{hemp}  R. Hempfling, Phys.~Lett.~B329 (1994) 222.
\bibitem{pq}  R.D. Peccei and H.R. Quinn, Phys.~Rev.~Lett.~38 (1977) 1440;
              R.D. Peccei and H.R. Quinn, Phys.~Rev.~D16 (1977) 1791.
\bibitem{kim} For reviews and more references see, J.E. Kim, Phys.~Rep.~150
             (1987) 1; H.-Y. Cheng, Phys.~Rep.~158 (1988) 1; R.D. Peccei, in
             {\it CP Violation}, ed.~C. Jarlskog (WSPC, Singapore, 1989) 503.
\bibitem{bbs} K.S. Babu, S.M. Barr and D. Seckel, preprint BA-94-21,
              hep-ph/9406308.
\bibitem{ckn} E.J. Chun, J.E. Kim and H.P. Nilles, Nucl.~Phys.~B370 (1992) 105;
              J.E. Kim and H.P. Nilles, preprint SNUTP 94-55, hep-ph/9406296.
\bibitem{gm} G.F. Guidice and A. Masiero, Phys.~Lett.~B206 (1988) 480;
             J.A. Casas and C. Munoz, Phys.~Lett.~B306 (1993) 288; A. Brignole,
            L.E. Ibanez and C. Munoz, Nucl.~Phys.~B422 (1994) 125;
             I. Antoniadis, E. Gava, K.S. Narain and T.R. Taylor, preprint
              IC/94/72 and hep-th/9405254.
\bibitem{zw} S. Ferrara, C. Kounnas, M. Porrati and F. Zwirner, Nucl.~Phys.
             B318 (1989) 75; S. Ferrara, C. Kounnas and F. Zwirner, Nucl.~Phys.
             B429 (1994) 589; A. Brignole and F. Zwirner, preprint
             CERN-TH.7439/94, hep-th/9409099 (to appear in Phys.~Lett. B).
\bibitem{mz} R.N. Mohapatra and X. Zhang, Phys.~Rev.~D49 (1994) R1163;
              E.J. Chun, H.B. Kim and A. Lukas, Phys.~Lett.~B328 (1994) 346.
\bibitem{dfsz} A.R. Zhitnitskii, Sov.~J.~Nucl.~Phys.~31 (1980) 103;
               M. Dine, W. Fischler and M. Srednicki, Phys.~Lett.~B104
               (1981) 199.
\bibitem{rtw} K. Rajagopal, M.S. Turner and F. Wilczek, Nucl.~Phys.~B358
              (1994) 447.
\bibitem{gy} T. Goto and M. Yamaguchi, Phys.~Lett.~B276 (1992) 103;
             E.J. Chun, J.E. Kim and H.P. Nilles, Phys.~Lett.~B287 (1992) 123.
\bibitem{ekn} J. Ellis, J. E. Kim and D. V. Nanopoulos,  Phys.~Lett.~B145
             (1984) 181; R. Juskiewicz, J. Silk and A. Stebbins, Phys.~Lett.
             B158 (1985) 463; J. Ellis, D. V. Nanopoulos and S. Sarkar,
             Nucl.~Phys. B219 (1985) 175; M. Kawasaki  and K. Sato,
             Phys.~Lett. B189 (1987) 23; T. Moroi, and H. Murayama and
             M. Yamaguchi,  Phys.~Lett.  B303 (1993) 289.
\end{thebibliography}
\end{document}